
%

\magnification=1200
\hsize 15true cm \hoffset=0.5true cm
\vsize 23true cm
\baselineskip=15pt
\nopagenumbers

\font\medio=cmr10 scaled \magstep2
\outer\def\beginsection#1\par{\medbreak\bigskip
      \message{#1}\leftline{\bf#1}\nobreak\medskip\vskip-\parskip
      \noindent}

\def\laq{\raise 0.4ex\hbox{$<$}\kern -0.8em\lower 0.62
ex\hbox{$\sim$}}
\def\gaq{\raise 0.4ex\hbox{$>$}\kern -0.7em\lower
0.62 ex\hbox{$\sim$}}

\def \at {\tilde \alpha}
\def \xt {\tilde x}

\def \pa {\partial}
\def \ra {\rightarrow}

\def \ti {\tilde}

\def \Da {\Delta}

\def \a {\alpha}

\def \Sg {\Sigma}
\def \sg {\sigma}
\def \da {\delta}

\def \r {\rho}

\def \noi {\noindent}

\def\sqr#1#2{{\vcenter{\hrule height.#2pt\hbox{\vrule width.#2pt
height#1pt \kern#1pt\vrule width.#2pt}\hrule height.#2pt}}}

\def\lsim{\mathrel{\rlap{\lower4pt\hbox{\hskip1pt$\sim$}}
    \raise1pt\hbox{$<$}}}         
\def\gsim{\mathrel{\rlap{\lower4pt\hbox{\hskip1pt$\sim$}}
    \raise1pt\hbox{$>$}}}         

\nopagenumbers
\line{\hfil CERN-TH/95-37}
\line{\hfil DFTT-14/95}
\vskip 1.5 true cm
\centerline {\medio VON NEUMANN AND SHANNON-WEHRL ENTROPY}
\vskip 0.5 true cm
\centerline{\medio FOR SQUEEZED STATES AND}
\vskip 0.5 true cm
\centerline{\medio COSMOLOGICAL PARTICLE PRODUCTION}

\vskip 0.8true cm
\centerline{M. Gasperini}
\centerline{\it Dipartimento di Fisica Teorica,
Via P.Giuria 1, 10125 Turin, Italy}
\centerline{and}
\centerline{M. Giovannini}
\centerline{\it Theory Division, CERN, CH-1211, Geneva 23, Switzerland}
\vskip 1 true cm
\centerline{\medio Abstract}

\noindent
We show that the effective coarse graining of a two-mode
squeezed density matrix, implicit in the Wehrl approaches to a
semiclassical phase-space distribution, leads to results in
agreement with previous different definitions  of
entropy for the process of pair production from the vacuum. We
also present, in this context, a possible interpretation of the
entropy growth as an amplification (due to the squeezing) of our
lack of knowledge about the initial conditions, which gives rise
to an effective decoherence of the squeezed density matrix.

\vskip 1.5true cm
\centerline{---------------------------}
\centerline {To appear in the {\bf ``String gravity and physics at the
Planck energy scale "} }
\centerline { (World Scientific, Singapore, 1995)}

\vskip 1 true cm
\noindent
CERN-TH/95-37

\noindent
February 1995
\vfill\eject

\footline={\hss\rm\folio\hss}
\pageno=1

\centerline{\bf VON NEUMANN AND SHANNON-WEHRL ENTROPY}
\centerline{\bf FOR SQUEEZED STATES}
\centerline{\bf AND COSMOLOGICAL PARTICLE
PRODUCTION}
\bigskip
\centerline{M. Gasperini and M. Giovannini}
\centerline{\it Dipartimento di Fisica Teorica, Universit\'a di Torino,}
\centerline{\it Via P.Giuria 1, 10125 Turin, Italy}

\bigskip
\centerline{ABSTRACT}
\midinsert
\narrower
\noi
We show that the effective coarse graining of a two-mode
squeezed density matrix, implicit in the Wehrl approaches to a
semiclassical phase-space distribution, leads to results in
agreement with previous different definitions  of
entropy for the process of pair production from the vacuum. We
also present, in this context, a possible interpretation of the
entropy growth as an amplification (due to the squeezing) of our
lack of knowledge about the initial conditions, which gives rise
to an effective decoherence of the squeezed density matrix.

\endinsert
\vskip 1 cm
\noi
Some years ago it was suggested that the in quantum process of
pair production from the vacuum, induced by the action of a
time-varying classical background, the growth of the average
number of quanta of a given field should be naturally associated
to the growth of entropy of that field [1,2]. More recently, such an
entropy growth was quantified on the grounds of a squeezed
state approach to the process of particle production [3-7]. With
this approach one can easily compute, in particular, the total
entropy $S$ stored in the cosmological perturbations amplified
by inflation, and one obtains [4,6]
$$
S\simeq \left(H_f\over M_p\right)^{3/2} S_{cmb}\eqno(1)
$$
where $S_{cmb}$ is the thermal black-body entropy of the $3
K^o$ photon microwave background, $H_f$ the final curvature
scale of inflation, and $M_p$ the Planck mass.

The use of the squeezed state formalism to compute entropy
production might seem to give rise to a puzzle, since the
squeezed evolution is unitary. The solution of the puzzle is that,
of course, there is no loss of information {\it in principle}
associated to the evolution of the initial state into a final
squeezed quantum state (and so no possible information
paradox, like in the case of black-hole dynamics). The loss of
information occurs however {\it in practice}, as a consequence of
the process of measure of the observables characterizing the
final squeezed state. There are indeed variables (related to the
so-called ``superfluctuant" operators [8]) which are amplified by
the squeezed evolution and become thus available for
macroscopic observations, while other variables (the
``subfluctuant" ones [8]) are correspondingly ``squeezed" and
become unobservable.

A simple and intuitive way to illustrate this situation is to
consider a semiclassical description of the squeezing process as
a parametric amplification of the wave function, ``hitting" the
effective potential barrier of a Schroedinger-like equation (with
time-like  variable) [9,10]. The final amplitude of the wave
scattered by the barrier may be larger or smaller than the initial
amplitude, depending on the phase with which the wave enters
the barrier. What is macroscopically detected, however, is the
average amplification of the wave, out of an initial random
distribution of phases. This point was clearly stressed already in
the first paper on the cosmological amplification of tensor
perturbations [11]. A macroscopic observation of the quantum
fluctuations of the metric, amplified by the cosmic evolution,
traces out over phases.

In order to compute the entropy associated to a squeezing
process, this macroscopic loss of information has to be
taken into account in the form of a suitable ``coarse graining" of
the density matrix of the final (pure) squeezed state. Such a
coarse graining was originally performed in a Fock basis, i.e.
with respect to the eigenstates of the number operator, by
averaging over phases [3-5]. The number operator, however, is
only a particular example of superfluctuant variable which grows
in the case of a pair production process, while the conjugate
variable (the phase) is correspondingly squeezed. This suggests
the use, in general, of the eigenstates of the superfluctuant
operator as the natural basis for the coarse graining of a
squeezed density matrix [6,7]. A possible different basis, that of
coherent states with real amplitude, has been suggested
however in [12].

It is important to stress that all these different approaches
[3-7,12] to the entropy of a squeezed state exactly lead to
the same result when the entropy growth $\Da S_k$, for each
mode $k$, is evaluated in the large squeezing regime, namely
$$
\Da S_k\simeq 2r_k, \,\,\,\,\,\,\,\,\,\,\,\,\,\,\,\,\,\,\,\,
r_k >>1 \eqno(2)
$$
where $r_k$ is the modulus of the (complex) squeezing
parameter [8]. In the previously quoted papers, however, the
decoherence process which leads to a reduced density matrix $
\r_{red}$ (from which to compute the quantum Von Neumann
entropy) is imposed ``by hand", by setting the off-diagonal terms
of the matrix to zero in some chosen basis; at best, like in the
case of the number basis, it is justified in a semiclassical way
using a stochastic approach and a sort of random-phase
approximation [3,4]. The problem is that it is formally impossible
in the number basis, as well as in the more general
superfluctuant basis, to compute the quantum von Neumann
entropy ($-Tr \r_{red} \ln \r_{red}$) from a reduced matrix
($\r_{red}\not= \r_{red}^2$) obtained by tracing out phases,
or another subfluctuant variable. Indeed, a subfluctuant operator
does not commute with the conjugate superfluctuant one, so
that both operators cannot be simultaneously included in the
complete set of observables characterizing a pure squeezed
state.

In view of these preliminary remarks, we can say that the aim of
this paper is twofold. First of all we want to compute the
semiclassical Shannon-Wehrl entropy [13,14] for a two-mode
squeezed vacuum state, in order to show that also in that
approach one recovers the result (2) for the large squeezing
limit. This confirms that the entropy of a highly squeezed state is
very robust to the particular scheme of coarse graining
implemented (as discussed also in [15,16]).

In the second place we want to present a possible justification
to the reduction of the density matrix in a
superfluctuant basis [6,7], by interpreting such a reduction as the
consequence of our lack of knowledge about the initial
configuration which is the starting point of the squeezed
evolution. To this purpose we shall represent the initial state, in
general, as a statistical mixture of coherent states. This
representation automatically defines a quantum von Neumann
entropy associated to the initial mixture. This entropy however is
constant, as the squeezed evolution is unitary, and then it does
not contribute to the difference $\Da S$ between final and initial
entropy. However, if the statistical distribution satisfies (at least
approximately) a principle of equipartition of the probabilities
among the states of the mixture, then the off-diagonal terms  of
the density matrix turn out to be suppressed. An effective
decoherence is thus produced, whose associated entropy in the
superfluctuant basis is eventually amplified by the squeezed
evolution.

The arguments presented in this paper are intended to apply to
any process of pair production, described by a Bogoliubov
transformation connecting the particle $\{b,b^\dagger\}$ and
anti-particle $\{\ti b, \ti b^\dagger\}$ annihilation and creation
$|in \rangle$ operators to the $|out\rangle $ ones,
$\{ a,a^\dagger,\ti a, \ti a^\dagger\}$ . Such a transformation
reads, for each mode $k$,
$$
a_k = c_+(k)b_k+c_-^*(k)\ti b_{-k}^\dagger
$$
$$
\ti a_{-k}^\dagger = c_-(k)b_k+c_+^*(k)\ti b_{-k}^\dagger
\eqno(3)
$$
where the Bogoliubov coefficients $c_{\pm}(k)$ satisfy
$|c_+|^2-|c_-|^2=1$. By putting
$$
c_+(k)= \cosh r_k , \,\,\,\,\,\,\,\,\,\,\,\,\,\,
c_-^*(k)= e^{2i\theta_k} \sinh r_k \eqno(4)
$$
the transformation (3) can be rewritten as a unitary
transformation,
$$
a_k=\Sigma_k b_k \Sigma_k^\dagger,\,\,\,\,\,\,\,\,\,\,\,\,\,\,\,\,
\ti a_{-k}^\dagger=\Sigma_k \ti b_{-k}^\dagger
\Sigma_k^\dagger \eqno(5)
$$
generated by the two-mode squeezing operator [8]
$$
\Sigma_k= \exp \left( z^*_kb_k\ti b_{-k} -
 z_k b_k^\dagger \ti b_{-k}^\dagger \right), \,\,\,\,\,\,\,\,
z_k=r_ke^{2i\theta_k} \eqno(6)
$$
Our discussion can thus be applied in general to any dynamic
situation in which the initial vacuum state $|0\rangle$ evolves
towards a final two-mode squeezed vacuum state, $|z\rangle
=\Sg_k |0\rangle $, characterized by a non-vanishing value of the
squeezing parameter $r_k=|z_k|$ (related to the expectation
value of the number of produced particles by $\langle n_k\rangle
=\langle z_k|b^\dagger_k b_k|z_k\rangle=
\langle 0|a^\dagger_k a_k|0\rangle=
|c_-(k)|^2=\sinh^2 r_k$).

It is important to note that in the context of such a dynamic
evolution one can always define two operators $x$ and $\xt$,
called superfluctuant, whose variances $(\Da x)_z$, $(\Da \xt)_z$
are amplified with respect to their vacuum value
$(\Da x)_0$, $(\Da \xt)_0$, namely [8]
$$
(\Da x)_0 \ra (\Da x)_z =(\Da x)_0 e^r, \,\,\,\,\,\,\,\,\,
(\Da\xt)_0 \ra (\Da \xt)_z =(\Da \xt)_0 e^r \eqno(7)
$$
where  $(\Da x)_z^2=\langle z |x^2|z \rangle -
 (\langle z |x|z \rangle )^2$ (and the same for $\xt$). The
variance of the canonically conjugate operators $y, \ti y$ is
correspondingly squeezed, $(\Da y)_z =(\Da y)_0 e^{-r} =
(\Da \ti y)_z $. In terms of the superfluctuant variables the
operators $b$ and $b^\dagger$ have the differential
representation [6,7,8]
$$
b={i\over 2}e^{i\theta}(x-i\xt+\pa_x-i\pa_{\xt})
$$
$$
\ti b={i\over 2}e^{i\theta}(x+i\xt+\pa_x+i\pa_{\xt})\eqno(8)
$$
(henceforth the mode index $k$ is to be understood, if not
explicitly written, and any correlation among modes with
different $|k|$ will be neglected, following the coarse graining
approach of [2]). The normalized wave function $\psi_z$ for the
two-mode squeezed vacuum (such that $a\psi_z=0=\at \psi_z$)
becomes, in the $(x,\xt)$-space representation [6,7],
$$
\psi_z(x,\xt)=\langle x \xt|\Sg |0\rangle \equiv
\langle x \xt|z\rangle=\left(\sg \over \pi \right)^{1/2}
e^{-{\sg \over 2}(x^2+\xt^2)},
$$
$$
\sg= e^{-2r},\,\,\,\,\,\,\,\,\,\,\,\,\,\,  \langle z|z\rangle =1
\eqno(9) $$
The squeezed evolution $|0\rangle \ra |z\rangle= \Sg |0\rangle $
may thus represented, in the superfluctuant basis, as a scaling
transformation $x\ra \sqrt \sg x$, $\xt\ra \sqrt \sg \xt$, with
real positive parameter $\sg \leq 1$, related to the squeezing
parameter $r$ according to eq.(9).

In order to compute the Shannon-Wehrl entropy for a squeezed
state we recall that the classical entropy is defined, for any
given phase-space distribution $f(q,p)$, as
$$
S_{cl}(f)=-\int dq dp f(q,p)\ln f(q,p) \eqno(10)
$$
while the quantum (or von Neumann) entropy is defined in terms
of the density operator  $\r$ as
$$
S_q(\r)=-Tr \r \ln \r \eqno(11)
$$
For a pure state, in particular, $\r^2=\r$ and $S_q=0$. Suppose
now that we want to compute, for a given quantum state
represented by $\r$, its entropy in the ``semiclassical" limit. It is
well known that a possible procedure to obtain the semiclassical
limit of a quantum observable is to compute the expectation
value of that observable in the basis of the coherent states
$|\a\rangle $. According to Wehrl [14], a natural semiclassical
phase-space distribution associated to $\r$ is thus given by
$$
f(\a)={1\over \pi}\langle \a |\r|\a\rangle \eqno(12)
$$
(the so-called Glauber distribution function [17]), which leads to
define the semiclassical (Wehrl) entropy $S_W$ as
$$
S_W= S_{cl} \left(f(\a)\right)\eqno(13)
$$

For a two-mode squeezed state $|z\rangle $ (see [18,19] for the
one-mode case) we have, in particular,
$$
f(\a,\at)={1\over \pi^2}\langle\a\at|z\rangle \langle
z|\a\at\rangle \eqno(14)
$$
where the coherent states $|\a\at\rangle$ satisfy
$$
b|\a\at\rangle= \a |\a\at\rangle,\,\,\,\,\,\,\,\,\,\,\,\,
\ti b |\a\at\rangle = \at |\a\at\rangle \eqno(15)
$$
(the operators $b, \ti b$ are defined in eq.(8); their corresponding
eigenvalues are complex numbers, $\a=\a_1+i\a_2$,
$\at=\at_1+i\at_2$). By expanding in the superfluctuant basis
$|x \xt \rangle$ one easily obtains
$$
\langle x \xt|\a \at\rangle ={1\over \sqrt \pi}
\exp \left [-{1\over 2}(\Re k)^2
-{1\over 2}(\Re \ti k)^2+k x+\ti k \xt -{1\over
2}(x^2+\xt^2)\right],
$$
$$
\langle\a\at|\a \at\rangle =1 \eqno(16)
$$
where $\Re z$ denotes the real part of $z$, and
$$
k= -i e^{-i\theta} (\a +\at),\,\,\,\,\,\,\,\,\,\,\,\,\,\,\,\,
\ti k= - e^{-i\theta} (\at -\a)\eqno(17)
$$
Moreover
$$
f(\a, \at)={1\over \pi^2}\int dx d\xt dx' d\xt'
\psi_z(x \xt)
\psi_z^*(x '\xt')\langle \a \at|x\xt\rangle
\langle x'\xt' |\a \at\rangle =
$$
$$
={1\over \pi^2 \cosh^2 r}\exp \left [-\tanh r (\a \at e^{-2i\theta}
+\a^*\at^* e^{2i\theta}) -|\a|^2 -|\at|^2 \right]\eqno (18)
$$
With this Glauber distribution, the semiclassical Wehrl entropy
(13) for the two-mode squeezed vacuum is then
$$
S_W= -\int d^2\a d^2\at f(\a,\at)\ln f(\a,\at) =
2+ 2\ln \pi +2\ln \cosh r \eqno(19)
$$
The same result holds for a squeezed-coherent state $|z \a_0
\at_0\rangle= \Sg |\a_0
\at_0\rangle $, obtained by applying a squeezing transformation
to the ``displaced" vacuum $|\a_0 \at_0\rangle = D(\a_0)D(\at_0)
|0\rangle$, where $D(\a)= \exp (\a b^\dagger - \a^* b)$ is the
Glauber displacement operator.

It is important to note that for $r>>1$ this semiclassical entropy
reproduces the large squeezing behavior of eq.(2), previously
obtained [1,3-7,12] with different approaches to the entropy of a
particle production process. We also note that the entropy (19) is
exactly the sum of two ``marginal" entropies,
$$
S_W= S^W_{\a_1\at _1} +S^W_{\a_2\at _2}
\eqno(20)
$$
obtained by decomposing the probability distribution $f(\a,\at)$
in the complex $\a,\at$ planes, and by tracing out one of the two
cartesian variables,
$$
S^W_{\a_1\at _1}= -\int d\a_1 d\at_1 f(\a_1, \at_1)\ln
f(\a_1, \at_1), \,\,\,\,\,\,
f(\a_1, \at_1)=\int d \a_2 d\at_2 f(\a, \at)
$$
$$
S^W_{\a_2\at _2}= -\int d\a_2 d\at_2 f(\a_2, \at_2)\ln
f(\a_2, \at_2), \,\,\,\,\,\,
f(\a_2, \at_2)=\int d \a_1 d\at_1 f(\a, \at)\eqno(21)
$$
This is a non-trivial result, as it implies that the
information-theoretical Araki-Lieb inequality [20],
$ S\leq S_{\a_1\at _1}+S_{\a_2\at _2}$, is maximized both by the
two-mode squeezed vacuum and by a squeezed-coherent state, in
agreement with the properties of one-mode squeezed states
[18].

According to eq.(19), the semiclassical entropy associated to a
pure squeezed state $|z\rangle$ is thus non-vanishing, in spite of
the fact that $\r=|z\rangle \langle z|=\r^2$ and $S_q(\r)=0$. This
occurs because an effective reduction of the quantum density
matrix is implicit in the semiclassical limit, as even a pure state
is seen, classically, as a disordered one. However, the coarse
graining associated to the semiclassical limit cannot give us
explicit information about the physical origin of the decoherence
mechanism, which can only be described in the context of some
dynamic model for the quantum to classical transition (see also
[21]). For a
squeezed state such a model must account, in particular, for the
fact that some degree of freedom of the system is made
dynamically less relevant than others for what concerns
macroscopic observations.

Let us consider now the possibility of defining also a von
Neumann entropy for the squeezing process, in terms of a
properly reduced density operator. In the approach of Refs.[5-7]
the reduction was performed by neglecting the off-diagonal
elements of the density matrix in a superfluctuant
representation. The use of such a representation is justified
because the squeezed density operator can only be reconstructed,
observationally, by measuring superfluctuant variables and their
momenta. But is the reduction a formally justified
procedure?

The answer is certainly not for the pure squeezed vacuum state
(9), whose density matrix in the superfluctuant representation
$$
\r_z(x,x',\xt, \xt')=\langle x \xt|z\rangle \langle z| x'\xt'\rangle
=\psi_z(x,\xt)\psi_z^*(x',\xt')=
$$
$$
={\sg\over \pi} \exp
\left [-{\sg \over 2}(x^2+x'^2+\xt^2+\xt'^2)\right] \eqno(22)
$$
is perfectly symmetric in the $(x,x')$ and $(\xt,\xt')$ planes. The
same answer holds for a pure squeezed number and
squeezed-coherent state. On the contrary, the reduction would
be naturally justified if the off-diagonal matrix elements would
be suppressed, $|\r_z(x,x)|>>|\r_z(x,x'\not= x)|$ (and the same
for $\xt$), for any given value of the squeezing parameter. It is
thus interesting to note that such requirement can be satisfied when
the initial state is not pure, but is represented by a
mixture, whose statistical weights tend to be equally distributed
among all the states of the mixture.

Suppose to start, in fact, from an initial configuration more
general than the vacuum, represented by an ensemble of states
which, like the vacuum, minimize however the quantum
fluctuations $\Da x$, $\Da\xt$ ( and their conjugate
$\Da y$, $\Da\ti y$). This corresponds to an initial density
operator which can be expressed in terms of the coherent states
(15), (16) as
$$
\r_i=\int d^2\a d^2\at P(\a,\at)|\a\at\rangle \langle
\a\at|\eqno (23)
$$
where the statistical weights $P\geq 0$ satisfy the
normalization condition
$$
\int d^2\a d^2\at P(\a,\at) =1\eqno(24)
$$
The squeezed evolution leads then to a final density operator,
$\r_f=\Sg \r_iÊ\Sg^\dagger $, which in the superfluctuant basis
is explicitly represented by
$$
\r_f(x,x',\xt, \xt')= \int d^2\a d^2\at P(\a,\at)
\psi_{z \a\at}(x,\xt) \psi_{z \a\at}^*(x',\xt ')
\eqno(25)
$$
Here $\psi_{z \a\at}(x,\xt)$ is the squeezed-coherent wave
function, representing in the ($x,\xt$) space the eigenfunctions
of $\{a, \ti aÊ\}$ with eigenvalues $\{\a, \at \}$, namely
$$
\psi_{z \a\at}(x,\xt) =
\langle x \xt|z \a\at\rangle  =\langle x \xt|\Sg| \a\at\rangle =
$$
$$
=\sqrt{\sg\over \pi} \exp
\left [-{1\over 2}(\Re k)^2 -{1\over 2}(\Re \ti k)^2 +\sqrt \sg
(k \sg +\ti k \xt)
-{\sg \over 2}(x^2+\xt^2)\right] \eqno(26)
$$
($k, \ti k$ are defined in eq.(17)). Note that, for $\r_i \not=
\r_i^2$, there is a  non-vanishing quantum entropy, $S_q= -
\int d^2\a d^2\at P(\a,\at) \ln P(\a\at)$, naturally associated to
the initial mixture. Such an entropy, however, is constant
throughout the whole process of squeezing $\r_i\ra \r_f$ (which
describes a unitary evolution), so that it cannot contribute to the
overall entropy difference $\Da S= S_f=S_i$.

The choice of the vacuum as initial condition corresponds to the
assumption $P(\a,\at)= \da^2(\a) \da^2(\at)$ in eq.(25). We shall
consider here an initial distribution which, instead of being
infinitely peaked upon the vacuum, tends to be spread rather
uniformly over a wide range of states, so as to reflect our lack
of knowledge about the initial conditions. We shall assume,
however, that such a distribution is always centered around the
vacuum, and we shall conveniently translate the ($x,\xt$) frame
in such a way that $\langle z \a \at|x|z \a \at\rangle = 0 =
\langle z \a \at|\xt|z \a \at\rangle $ (which implies $\Re
k=0=\Re \ti k$, see eq.(26)). The simplest example satisfying the
above requirements is a normalized square-like distribution, in
which the probability of a state $|\a\at\rangle$ is constant and
equal to $1/L$ for $|\a|, |\at|<L/2$, and vanishing for
$|\a|, |\at|>L/2$, namely
$$
\int d^2\a d^2\at P(\a,\at)=
d\a_1d\a_2d\at_1d\at_2
\da(\a_1-\at_1)\da(\a_2+\at_2) \times
$$
$$
\times
{1\over L^2}\left[\theta(\a_1+L/2)-\theta (\a_1-L/2)\right]
\left[\theta(\a_2+L/2)-\theta (\a_2-L/2)\right]
\eqno(27)
$$
Here $\theta$ is the Heaviside step function, and the two delta
distributions have been inserted to implement the conditions
$\langle x\rangle =0=\langle\xt\rangle$; we have also rescaled
the variables $\a,\at$ so as to absorb the overall phase
$e^{-i\theta}$ which does not contribute to the integration
over the whole complex plane.

With this probability distribution the squeezed matrix (25)
becomes
$$
\r_f(x,x',\xt, \xt')
={\sg\over \pi} {\sin \left [ L\sqrt \sg (x'-x)\right]\over
L\sqrt \sg (x'-x)}
{\sin \left [ L\sqrt \sg (\xt-\xt')\right]\over
L\sqrt \sg (\xt-\xt')}e^{
-{\sg \over 2}(x^2+x'^2+\xt^2+\xt'^2)} \eqno(28)
$$
and it is immediately evident that, for large enough $L$, the
off-diagonal terms $x\not=x'$, $\xt \not= \xt'$ are suppressed
with respect to the diagonal ones. Such a suppression grows with
$L$, namely with the number of states which are considered
equiprobable in the initial distribution, i.e. with the growth of our
``ignorance" about the initial configuration. For large $L$ we may
thus approximate the squeezed matrix with the diagonal
elements only, and in particular we can formally define in this
basis a (normalized) reduced density matrix, in the limit $L\ra
\infty$ (at finite $\sg$), as
$$
\r_{red}(x,x',\xt, \xt')=
\lim_{L \ra \infty} \{L^2 \sg
\r_f(x,x',\xt, \xt')\}=
$$
$$
={\sg\over \pi} e^{
-{\sg }(x^2++\xt^2)} \da (x-x') \da (\xt-\xt')
\eqno(29)
$$
This is exactly the representation in the ($x,\xt$) basis of the
reduced operator
$$
\r_{red}=
\int dx d\xt |\psi_z(x, \xt)|^2 |x \xt\rangle \langle x \xt|
\eqno(30)
$$
previously used in Refs.[6,7].

Note that $Tr \r_{red}=1$, but $\r_{red}^2\not=\r_{red}$. This
reduction leads then to a quantum entropy $S_q(\r_{red})$ which,
unlike the entropy of the mixture $S_q(\r_f)$, depends on $\sg$,
and in particular grows with the degree of squeezing. By taking
the difference between final and initial entropy we find indeed
the result of eq.(2),
$$
\Da S= \left( Tr \r_{red} \ln \r_{red}\right)_{\sg=1}-
\left( Tr \r_{red} \ln \r_{red}\right)_{\sg}= 2r \eqno(31)
$$
The accuracy of this estimate in our context becomes larger and
larger as the initial probability distribution becomes flatter, so
as to share probabilities among a larger number of states. It is
also important to note that if, for a given $L$, the reduced matrix
represent a good approximation to the exact matrix (28) at large
squeezing ($\sg>>1$), then the approximation is even better as
we go back in time towards the initial configuration, where
$\sg=1$.

We may thus conclude that the effective reduction of the
squeezed density operator, introduced ``ad hoc" in previous
papers [6,7], can be formally justified through  a limiting
procedure based on an appropriate mixture of coherent states,
assumed to represent the initial configuration. The corresponding
entropy growth may thus physically interpreted as an
amplification, due to the squeezed evolution, of our ignorance
about the initial conditions, in a context in which we have no
motivation to single out some preferred initial state, and we
cannot reconstruct the initial state through final measurements
of superfluctuant variables only. Such an interpretation seems to
be particularly appropriate in the context of cosmological
perturbation theory, where the amplification of the background
fluctuations is computed starting from an initial state which is
left unspecified except for the requirement of minimizing the
quantum fluctuations, and which can thus be properly
represented as a general mixture of coherent states.
\vskip 2 cm
{\bf Acknowledgments}

One of us (M. Gasperini) wishes to thank D. Boyanovsky for rising
stimulating questions that motivated in part the work reported
here. We are also  grateful to G. Veneziano for many helpful
discussions.
\vskip 2cm
\centerline{\bf References}

\item{1.}B. L. Hu and D. Pavon, Phys. Lett. B 180, 329 (1986)

\item{2.}B. L. Hu and H. E. Kandrup, Phys. Rev. D 35, 1776 (1987);

H. E. Kandrup, Phys. Rev. D  37, 3505 (1988);

for a recent review on cosmological particle production see also
B. L. Hu,

G. Kang and A. Matacz, Int. J. Mod. Phys. A 9, 991 (1994)

\item{3.}R. Brandenberger, V. Mukhanov and T. Prokopec, Phys.
Rev. Lett. 69, 3606 (1992)

\item{4.}R. Brandenberger, V. Mukhanov and T. Prokopec, Phys.
Rev. D 48, 2443 (1993)

\item{5.}T. Prokopec, Class. Quantum Grav. 10, 2295 (1993)

\item{6.}M. Gasperini and M. Giovannini, Phys. Lett. B 301, 334
(1993)

\item{7.}M. Gasperini and M. Giovannini, Class. Quantum Grav. 10,
L133 (1993)

\item{8.}B. L. Shumaker, Phys. Rep. 135, 317 (1986);

J. Grochmalicki and M. Lewenstein, Phys. Rep. 208, 189 (1991)

\item{9.}B. L. Hu, Phys. Rev. D 9, 3263 (1974)

\item{10.}L. P. Grishchuk and Y. V. Sidorov, Phys. Rev. D 42, 3413
(1990);

L. P. Grishchuk and L. Solokhin, Phys. Rev. D 43, 2566 (1991)

\item{11.}L. P. Grishchuk, Zh. Eksp. Teor. Fiz. 67, 825 (1975)

\item{12.}A. Matacz, Phys. Rev. D 49, 788 (1994)

\item{13.}C. E. Shannon and W. Weaver, ``The mathematical theory
of communication" (Univ. of Illinois Press, Urbana, IL, 1963)

\item{14.}A. Wehrl, Rev. Mod. Phys. 50, 221 (1978)

\item{15.}E. Keski-Vakkuri, Phys. Rev. D49, 2122, (1994)

\item{16.}M. Kruczenski, L. E. Oxman and M. Zaldarriaga, ``Large
squeezing behavior of cosmological entropy generation",
Preprint gr-qc/9403024

\item{17.}R. London and P. L. Knight, J. Mod. Opt. 34, 709 (1987)

\item{18.}C. Keitel and K. Wodkiewicz, Phys. Lett. A 167, 151
(1992)

\item{19.}H. Rosu and M. Reyes, ``Shannon-Wehrl entropy for
cosmological and black-hole squeezing", Preprint gr-qc/9406001

\item{20.}H. Araki and E. Lieb, Commun. Math. Phys. 18, 160 (1970)

\item{21.}W. H. Zurek, S. Habib and J. P. Paz, Phys. Rev. Lett. 70,
1187 (1993)

\end